# Deep learning assisted inverse design of nonreciprocal multilayer photonic structures

Weiran Zhang[1], Hao Pan[1], and Shubo Wang[1,*]

[1] Department of Physics, City University of Hong Kong, Tat Chee Avenue, Kowloon, Hong Kong, China

*Corresponding author: shubwang@cityu.edu.hk

**Abstract:** Nonreciprocal structures play an important role in optical physics and applications. Conventional approaches for designing nonreciprocal optical structures rely heavily on extensive numerical simulation and parameter tuning, leading to high computational cost and low efficiency. Here, we apply deep learning to the design of nonreciprocal multilayer photonic structures. Three neural-network models—a forward neural network (FNN), an inverse design network (IDN), and a variational autoencoder (VAE)—are employed to learn the complex mapping between structural/material parameters and nonreciprocal spectral characteristics. We show that the FNN can rapidly and accurately predict the nonreciprocal electromagnetic response of a given structure, while the IDN can directly generate suitable structural parameters for target spectral responses. Both approaches substantially reduce computational cost and design time while improving nonreciprocal performance. Furthermore, the VAE can generate band-limited inverse design under practical performance constraints, facilitating efficient exploration of multiple feasible structures that meet different threshold requirements within specified frequency bands. Our work highlights the potential of deep learning for the advanced design of nonreciprocal optical structures and devices.

# 1. Introduction

Nonreciprocal systems, which break the constraints of Lorentz reciprocity, are essential for modern photonic technologies, playing a critical role in the design of optical isolators, circulators, and nonreciprocal meta-devices[1-6]. While nonreciprocal electromagnetic responses can be realized through magnetic[7], nonlinear[8], moving[9,10], or time-varying materials[11], the Faraday magneto-optical effect in magnetic materials remains a fundamental and widely used mechanism for breaking time-reversal symmetry and achieving robust nonreciprocity[12-14].

Conventional approaches for designing nonreciprocal photonic structures rely extensively on complex modeling and iterative parameter tuning. These approaches demand substantial computational resources and often lack design flexibility, making the development of straightforward and efficient alternatives a priority[15-17]. Data-driven approaches based on machine learning have recently emerged as a promising solution. The use of machine learning in photonics dates back to the second wave of artificial intelligence research[18-20], when multilayer perceptrons were employed as computer-aided design tools for microwave or radio-frequency circuits[21,22]. More recently, deep learning has established itself as a dominant approach in AI-assisted photonic research[23,24]. Enabled by advances in training and regularization techniques[25-28], it is now feasible to train deeper neural networks that exhibit significantly improved performance[29-34].

Recent years have witnessed rapid methodological advances in AI-assisted photonic design. Modern frameworks now integrate forward and inverse mappings to simplify training[35], employ transfer learning to accelerate convergence for complex structures[36], and utilize optimization-oriented networks to handle multi-objective tasks[37]. Beyond these engineering capabilities, deep learning has also emerged as a powerful tool for exploring fundamental physics. Its success in broader physical systems, such as learning dynamics that respect conservation laws[38] and identifying phase transitions in condensed matter[39], highlights its immense potential for uncovering complex mechanisms in photonic systems as well.

Motivated by these developments, we apply deep neural networks to model and design nonreciprocal multilayer photonic structures. Specifically, a forward neural network (FNN) is trained to rapidly predict transmittance, reflectance, and absorptance spectra from structural/material parameters. Meanwhile, an inverse design network (IDN) is utilized to accurately retrieve structural/material parameters from target spectra. Both models demonstrate excellent agreements between the predicted and ground-truth values, verifying the effectiveness and reliability of our approach. Furthermore, by analyzing the prediction results in the context of underlying physics, we identify the relation between neural network performance and the physical properties of the structures. These results pave the way for future applications of deep learning in investigating and designing complex nonreciprocal optical systems.

The rest of this paper is organized as follows. Section 2 introduces the nonreciprocal photonic structure and the transfer matrix method (TMM) for generating the dataset. Sections 3 and 4 present the FNN and the IDN that map the relationships between design parameters and spectra. Section 5 presents the band-limited design of nonreciprocal responses using VAE model for optimization. Section 6 discusses the results and provides further analysis of the neural network performance and the AI-informed physical insights. Finally, Section 7 concludes the paper and outlines potential future directions.

## 2. Transfer Matrix Method

As illustrated in Fig. 1, we consider an incident electromagnetic plane wave propagating through a multilayer structure. The transfer matrix method (TMM) [40] can be employed to generate the dataset for training and validating the deep learning models. We first consider the *s*-polarization with the electric field perpendicular to the incident plane (*yoz*-plane). By enforcing the continuity of the electric and magnetic fields at the interface between regions 1 and 2 in Fig. 1, we obtain:

$$E_1 + E_1^r = E_2 + E_2^r$$
$$\sqrt{\frac{\varepsilon_1}{\mu_1}}\left(E_1 - E_1^r\right)\cos\theta_1 = \sqrt{\frac{\varepsilon_2}{\mu_2}}\left(E_2 - E_2^r\right)\cos\theta_2, \tag{1}$$

where the fields with (without) superscript "*r*" represents the forward (backward) propagating wave. These boundary conditions can be rewritten in matrix form by introducing the dynamical matrix as

$$D(1)\begin{pmatrix} E_1 \\ E_1^r \end{pmatrix} = D(2)\begin{pmatrix} E_2 \\ E_2^r \end{pmatrix} \tag{2}$$

where

$$D(i) = \begin{pmatrix} 1 & 1 \\ \sqrt{\frac{\varepsilon_i}{\mu_i}}\cos\theta_i & -\sqrt{\frac{\varepsilon_i}{\mu_i}}\cos\theta_i \end{pmatrix} \tag{3}$$

This formulation can be applied to all interface boundaries in the multilayer structure.

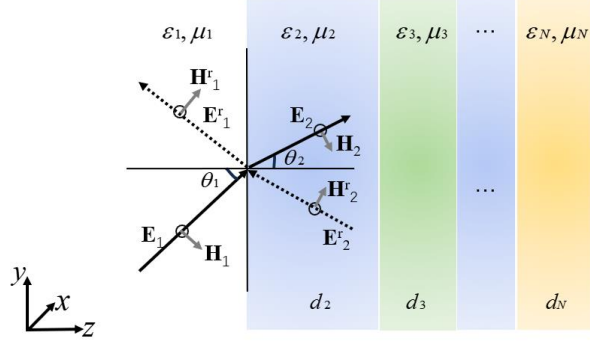

Fig.1. Schematic of electromagnetic wave propagation from air into a multilayer structure. The forward-propagating wave is represented by electric field $\mathbf{E}_i$ and magnetic field $\mathbf{H}_i$, while the reflected wave is denoted by $\mathbf{E}_i^r$ and $\mathbf{H}_i^r$. The first region corresponds to air. Each layer is characterized by the material parameters $\varepsilon_i$, $\mu_i$ and thicknesses $d_i$, where $i=1, 2, …, N$.

The phase accumulation within each individual layer can be characterized by a propagation matrix $P(i)$. Specifically, as the wave propagates through the $i$-th layer with thickness $d_i$, the forward wave acquires a phase factor $e^{ik_{zi}d_i}$ while the backward wave acquires a phase factor $e^{-ik_{zi}d_i}$, where $k_{zi}$ is the wavevector in the $i$-th layer. Thus, the propagation matrix can be expressed as

$$P(i) = \begin{pmatrix} e^{ik_{zi}d_i} & 0 \\ 0 & e^{-ik_{zi}d_i} \end{pmatrix} \qquad (4)$$

in the same basis with the dynamical matrix.

To construct the total transfer matrix of the multilayer system, we note that the wave first encounters an interface described by $D(i)$, then propagates through a homogeneous layer described by $P(i)$, and finally exits the layer via another interface described by the inverse dynamical matrix $D(i)^{-1}$. This process is repeated for all intermediate layers, and the entire system is enclosed by the boundaries of the first and last media $D(1)$ and $D(N+1)$ respectively. Therefore, the incident and reflected fields can be expressed as

$$\begin{pmatrix} E_1 \\ E_1^r \end{pmatrix} = \begin{pmatrix} M_{11} & M_{12} \\ M_{21} & M_{22} \end{pmatrix} \begin{pmatrix} E_{N+1} \\ E_{N+1}^r \end{pmatrix} \qquad (5)$$

with the transfer matrix given by

$$\begin{pmatrix} M_{11} & M_{12} \\ M_{21} & M_{22} \end{pmatrix} = D(1)^{-1} \cdot [\prod_{i=2}^{N} D(i)P(i)D(i)^{-1}] \cdot D(N+1) \qquad (6)$$

The reflection and transmission coefficients can be directly computed from the above matrix elements. Assuming unit amplitude for the incident wave from the left and no incoming wave from the right (i.e., $E_{N+1}^r = 0$), the reflectance $R$ and transmittance $T$ can be defined as

$$R = \left|\frac{E_1^r}{E_1}\right|^2 = \left|\frac{M_{21}}{M_{11}}\right|^2 \qquad T = \left|\frac{E_{N+1}}{E_1}\right|^2 = \left|\frac{1}{M_{11}}\right|^2 \qquad (7)$$

provided that the wave is incident from air and eventually transmitted into air.

In our case, we consider the multilayer structure composed of alternating layers of YIG and dielectric materials under the normal incidence of circularly polarized plane waves. For normal incidence, the equations above for $p$-polarization and $s$-polarization reduce to the same form, and the analysis above can be directly applied to our nonreciprocal system. When biased by an external magnetic field in $+z$-direction, the permeability tensor of YIG can be expressed in the circular basis of $xy$-plane as

$$\mu = \mu_0 \begin{bmatrix} \mu_r & i\kappa_r & 0 \\ -i\kappa_r & \mu_r & 0 \\ 0 & 0 & \mu_r \end{bmatrix} \qquad (8)$$

where $\mu_r$ and $\kappa_r$ are frequency-dependent parameters defined as[41,42]:

$$\mu_r = 1 + \frac{\omega_m(\omega_0 + i\alpha\omega)}{(\omega_0 + i\alpha\omega)^2 - \omega^2}, \quad \kappa_r = \frac{\omega\omega_m}{(\omega_0 + i\alpha\omega)^2 - \omega^2} \qquad (9)$$

where $\omega_0 = \gamma M_0$ is the Larmor precession frequency; $\omega_m = \gamma M_s$ represents the strength of the magnetization; $\gamma = 1.76 \times 10^7$ rad s$^{-1}$ Oe$^{-1}$ is the gyromagnetic ratio; $M_s = 1800$ Oe is the saturation magnetization; $M_0 = 3570$ Oe is the external magnetic field; and $\alpha = 0.002$ is the damping constant. All quantities are expressed in Gaussian units. The off-diagonal term $\kappa_r$ accounts for the nonreciprocal properties of the system.

## 3. Forward neural network for predicting nonreciprocal responses

In this section, we demonstrate how to apply a FNN to map the structural parameters to optical responses. The structure is shown in Fig. 2(a), which consists of a five-layer stack with alternating layers of YIG and dielectric materials. Specifically, the structure follows the $\text{YIG} - \varepsilon_1 - \text{YIG} - \varepsilon_2 - \text{YIG}$ arrangement, forming a nonreciprocal system. The material properties of YIG are described by Eq. (9), and its geometric thicknesses $(d_1, d_3, d_5)$ are variable within the range of 2–6 mm. The permittivities of two dielectric layers, $\varepsilon_1$ and $\varepsilon_2$, are tunable between 2 to 10, covering the permittivity range of common dielectric materials. And their thicknesses $(d_2, d_4)$ are also variable within the range of 2–6 mm. Therefore, the input parameters for the FNN model are $[d_1, d_2, d_3, d_4, d_5, \varepsilon_1, \varepsilon_2]$. The two incidence directions in Fig. 2(a) represent a pair of circularly polarized plane waves related by time reversal symmetry, used to probe and verify nonreciprocal properties of the structure. The outputs of the FNN model are the spectral differences in transmittance, reflectance, and absorptance between these two excitations, sampled at 190 evenly spaced frequency points over the range of 1–20 GHz.

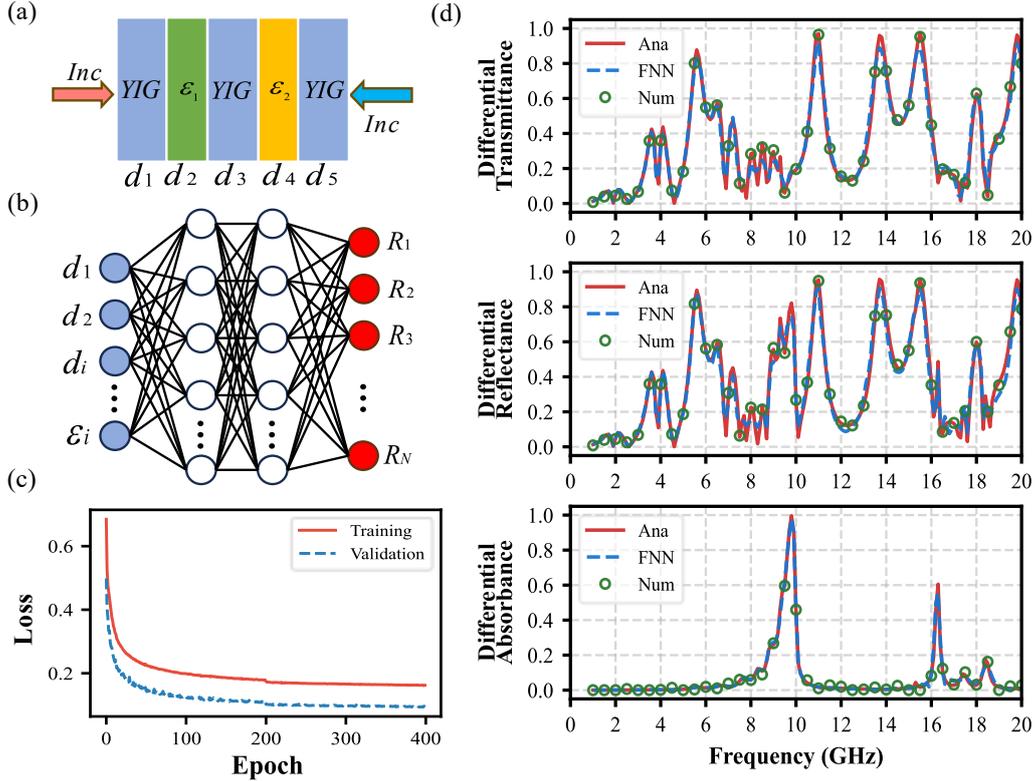

Fig. 2. (a) Schematic of a five-layer nonreciprocal stack model composed of alternating layers of YIG and dielectric materials. (b) Architecture of the FNN that can predict the spectra from structural thicknesses and material properties. (c) Training and validation loss over epochs. (d) Comparison of the spectra among analytical results (solid lines), FNN predictions (dashed lines), and numerical simulations (circles) for differential transmittance (top), differential reflectance (middle), and differential absorbance (bottom).

As shown in Fig. 2(b), we implement a fully connected multilayer perceptron neural network[43-45] to learn the complex nonlinear mappings between the system parameters and the nonreciprocal responses. We generate 50,000 samples of the dataset by using the TMM in Section 2, and they were randomly divided into training, validation, and test datasets in an 8: 1: 1 ratio. The architecture of the FNN is 7−256−512−512−384−128−190, where 7 and 190 are the number of inputs and outputs, and the remaining ones are the number of neurons in hidden layers. The loss function is the mean squared error. We also apply a learning rate annealing strategy to ensure stable convergence with its initial value of $10^{-3}$, and the learning rate is reduced by a factor of 0.75 if the validation loss stops improving for 10 consecutive epochs.

The training and validation loss over 400 epochs are shown in Fig. 2(c). We notice that both losses decrease rapidly during the first 50 epochs, followed by a gradual and stable convergence. This indicates that the model generalizes well and does not overfit the training data. Then, we apply this trained FNN to predict the differential transmittance, differential reflectance, and differential absorbance of the multilayer structure within the frequency range of [1 GHz, 20 GHz] under the opposite incidences. The parameters of this structure are: [2.59, 3.82, 5.24, 2.63, 2.70, 3.9, 4.1]. The results given by the FNN are shown in Fig. 2(d) as the dashed lines, while the analytical results computed with TMM are denoted by the solid lines, and the numerical simulations by COMSOL are represented by the circles. As seen, they exhibit a notable level of agreement for the three spectra, demonstrating the good performance of the FNN in predicting the nonreciprocal responses of the structures.

## 4. Inverse design network for retrieving structural parameters

We can also apply a deep learning neural network (i.e., IDN) to map the relationships from spectral responses to the design parameters. However, directly training the IDN often results in slow convergence or even instability. This challenge arises due to the non-uniqueness of the inverse mapping, as multiple distinct structural designs can give rise to similar spectral behavior, corresponding to the one-to-many problem.

To overcome this issue, we adopt a tandem neural network[35], as illustrated in Fig. 3(a). Instead of training the inverse model independently, we combine it with a pre-trained FNN, which remains fixed, and only the IDN parameters were updated. The IDN first predicts the design parameters from a given target spectrum, and these predicted parameters are then passed into the FNN to regenerate the spectral response. After these steps, we only need to minimize the loss between the target spectral response and the response reconstructed by the forward model, without requiring the real design parameters to be the same.

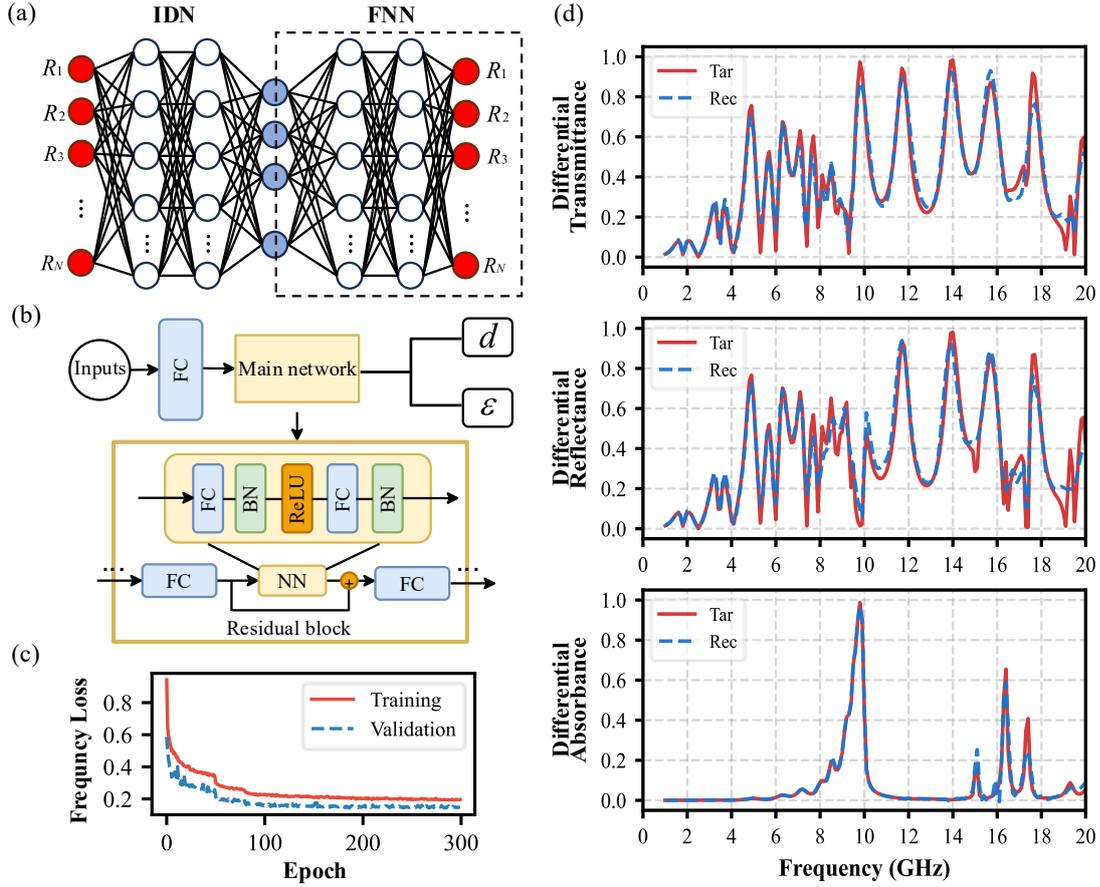

Fig. 3. (a) Architecture of the tandem neural network that combining an IDN and a pre-trained FNN. (b) Detailed structure of the IDN. The input passes through a fully connected (FC) layer and a series of residual blocks, each consisting of FC layers, batch normalization (BN), and ReLU activation functions. (c) Frequency loss over epochs of the IDN, described in Eq. (10). (d) Comparison of target (solid lines) and reconstructed (dashed lines) spectra for differential transmittance (top), differential reflectance (middle), and differential absorbance (bottom).

To improve the training process of the IDN, we incorporate residual blocks into a deeper network architecture, as shown in Fig. 3(b). Each residual block consists of fully connected layers with batch normalization and ReLU (rectified linear units) activation[25], along with skip connections that directly pass the input to the output. Residual networks were originally developed in the computer vision domain to enable stable training of very deep architectures[46]. During the early stages of training, the residual components dominate, allowing the network to behave as a deep architecture, which is beneficial for complex inverse mappings. As training progresses and the loss becomes smoother, the network begins to skip the deep residual components, causing the network to behave like a shallow one. This dynamic depth adaptation allows the network to simplify itself during the later stages, thereby effectively mitigating overfitting.

The architecture of the IDN consists of a shared feature extractor, a main network, and two separate fully connected branches for thickness and permittivity prediction. The input of the model is a 190-dimensional vector corresponding to the discrete frequency points of spectrum. The architecture of the shared feature extractor is 190−256−256. The thickness prediction branch consists of two fully connected layers with 256 neurons and a residual block with 256 neurons, followed by a 128 neurons fully connected layer and an output layer with 5 neurons. The permittivity prediction branch receives a 261-dimensional input formed by concatenating the 256-dimensional shared feature vector with the 5 values from thickness prediction branch. It then passes through a 128-neuron fully-connected layer and a 128-neuron residual block, followed by a 64-neuron fully-connected layer and an output layer with 2 neurons. All network layers use ReLU activation functions.

Moreover, we design a multistage adaptive loss function to constrain the generated physical structures within reasonable ranges. The total loss is computed as the sum of mean squared errors between predicted and target values for frequency response, thickness, and permittivity, defined as:

$$\mathcal{L}_{\text{total}} = \mathcal{L}_{\text{freq}} + \lambda \cdot \left( \mathcal{L}_{\text{thickness}} + \mathcal{L}_{\text{permittivity}} \right), \tag{10}$$

where $\lambda$ is the coefficient to adjust the weights of different loss parts. And we divide the total training process into four phases:

(1) Epochs (0–50), $\lambda = 1$: encourage the model to assign equal importance on spectrum and design parameter losses.

(2) Epochs (50–80), $\lambda = 0.5$: enhance the model to improve spectrum features.

(3) Epochs (80–120), $\lambda = 0.1$: emphasize spectral domain for accurate reconstruction.

(4) Epochs (120–300), $\lambda = 0.1$: maintain the weight but reduce learning rate by a factor of 0.75. The training and validation loss over 300 epochs are shown in Fig. 3(c). We see that the frequency response loss steadily decreases to a minimal value, confirming the model's ability.

We then apply the IDN to retrieve the structural parameter vector $[d_1, d_2, d_3, d_4, d_5, \varepsilon_1, \varepsilon_2]$ for a target differential transmittance spectrum (i.e., the transmittance differences between LCP and RCP incidences). These parameters are then fed into the forward model to regenerate the full spectral response. As shown in Fig. 3(d), the reconstructed spectra exhibit excellent agreement with the target spectra, demonstrating the reliability of the IDN in designing multilayer photonic structures with desired nonreciprocal properties.

## 5. Band-limited inverse design for nonreciprocal response

In the previous section, the IDN is trained to retrieve structural parameters from given target spectra. This approach is effective when a complete spectrum is given over the entire frequency range. However, from a practical perspective, such a requirement is often unnecessary. In most nonreciprocal devices, the design objective is usually not to reproduce a spectrum point by point, but to ensure that the nonreciprocal response remains above a threshold level within a given operational frequency band.

Under this band-limited requirement, the solution of the inverse problem becomes inherently non-unique, as multiple distinct structures may satisfy the same performance criterion. To efficiently search for structures through neural network, we employ a VAE[47] to model the distribution of valid structural parameters. The VAE is trained directly on the same dataset used for the forward and inverse networks in Sections 3 and 4. The encoder maps a structural parameter vector $p$ into a latent variable $z$ described by Gaussian distribution like $q(z|p) = N(\mu(p), diag(\sigma^2(p)))$, and the decoder reconstructs the parameters from the latent variable $z$. The loss function is composed of reconstruction loss together with a KL-divergence regularization term, defined as: $\mathcal{L}_{\text{VAE}} = \|p - p\|^2 + \beta D_{\text{KL}}\left(q_\phi(z|p) \| \mathcal{N}(\mathbf{0}, \mathbf{I})\right)$ [48], where $\beta$ is gradually increased during training to stabilize convergence.

After training, the VAE provides a compact representation of the parameter space learned from the dataset. New candidate structures can then be generated by sampling latent variables from the latent space. Compared with random sampling in the original parameter space, the VAE can generate candidate designs more efficiently by focusing on physically meaningful regions of the design space. For each generated structure, the corresponding nonreciprocal spectrum is then calculated directly from the pre-trained FNN introduced in Section 3.

To quantify the band-limited performance, we consider the minimum transmittance difference within the target frequency band defined as: $f(p) = \min \Delta T(f), f \in [f_1, f_2]$ satisfying $f(p) \geq \eta$. Both the target frequency range $[f_1, f_2]$ and the threshold $\eta$ can be flexibly adjusted according to design requirements. Using this criterion, candidate structures are generated by the VAE and evaluated by the FNN to check whether they satisfy the threshold condition, and the satisfied ones will be retained.

We consider an example with the target band of 12–14 GHz and a threshold of $\eta = 0.8$. We apply the VAE to generate the design parameters of the multilayer structure, and then apply the FNN to obtain the differential transmittance. The result is shown in Fig. 4(a). We notice that the designed structure gives rise to transmittance difference above 0.83 across the entire band of 12–14 GHz. As shown in Fig. 4(b), we further test a wider-band design by increasing the target band to 11–15 GHz. In this case, we find that no structure can be generated for the same threshold $\eta = 0.8$. This indicates that the performance requirement (i.e., a larger bandwidth) becomes overly demanding, which may exceed the physically achievable limit. Consequently, the VAE model cannot produce solutions. By relaxing the threshold to $\eta = 0.65$, a valid solution can be successfully obtained. The corresponding transmittance difference is shown in Fig. 4(b), where the difference is above 0.68 for the whole required bandwidth of 11–15 GHz. These examples demonstrate that our VAE model can efficiently generate designs of multilayer photonic structures with desired band-limited performance, which is particularly beneficial for practical applications requiring tunable specifications.

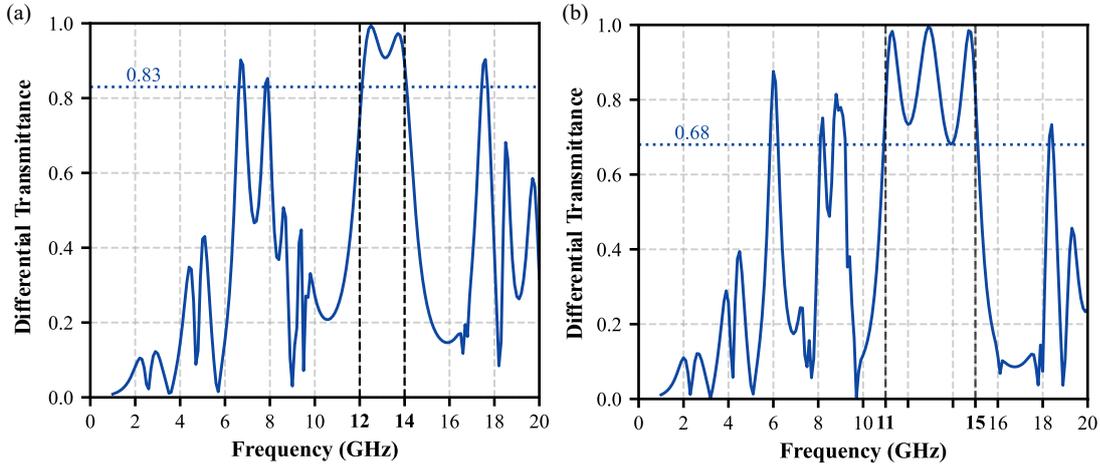

Fig. 4. (a) Band-limited differential transmittance of the VAE-designed structures with the target band of 12–14 GHz and a threshold of $\eta = 0.8$. (b) Band-limited differential transmittance of the VAE-designed structures with the target band of 11–15 GHz and a threshold of $\eta = 0.65$. The black dotted lines indicate the target frequency bands, and the blue dotted lines denote the minimum values over target bands.

## 6. Performance analysis

We now evaluate the performance of the neural network models from a statistical perspective using the coefficient of determination, i.e., the R² value, which is defined as[49]:

$$R^2 = 1 - \frac{\sum_{i=1}^{N}(y_i - \hat{y}_i)^2}{\sum_{i=1}^{N}(y_i - \bar{y})^2} \qquad (11)$$

where $y_i$ and $\hat{y}_i$ denote the ground truth values and the predicted values, respectively; $\bar{y}$ is the mean of the ground truth values. This quantifies the proportion of variance in the ground truth values that can be explained by the model's predictions, with value closer to one indicating better predictive accuracy.

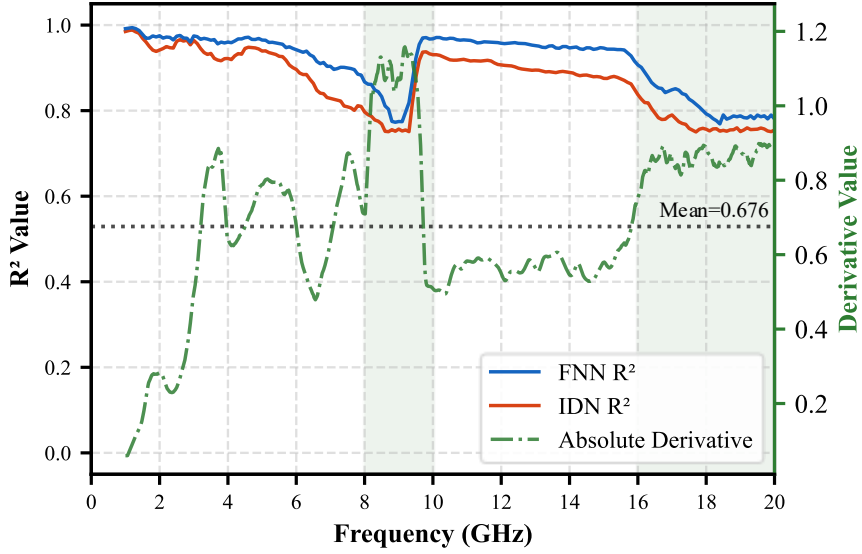

Fig. 5. R² distributions for the FNN (blue solid line) and IDN (red solid line), along with the absolute spectral derivative (green dashed line, right axis) as a measure of spectral complexity. Regions with large derivatives (highlighted in green) correspond to lower R² values, indicating that rapid spectral variations impose greater challenges for both prediction and inverse design. The dashed horizontal line denotes the global mean derivative value.

The R² distribution across frequency spectrum of the FNN and IDN are shown in Fig. 5 as the solid blue and red lines, respectively. The total $R^2$ values are 0.90 in the FNN and 0.86 in the IDN, both indicating a strong agreement between the predicted and actual values. From the figure, we observe a similar trend that the prediction accuracy, as measured by R² value, tends to drop significantly at certain frequency bands while remaining consistently high at others. This frequency-dependent behavior of the performance can be attributed to the intrinsic complexity of the system at different frequency bands. Specifically, the regions with rapid variations in the spectrum may induce greater challenges for the deep learning models, which originates from the dispersive nature of the permeability of YIG material[50]. Using Eq. (8) and Eq. (9), the permeability for the RCP wave, given by the corresponding eigenvalue of the permeability tensor (assuming no attenuation, α = 0), is expressed as $\mu_0\left(\dfrac{\omega_m + \omega_0 - \omega}{\omega_0 - \omega}\right)$. This permeability exhibits a resonance at the frequency about 10.0 GHz, where the response of the system changes dramatically. In addition, when $f \to f_0 + f_m \approx 15.8 \text{GHz}$, the permeability approaches zero, which also can give rise to unusual responses of the system[51-53]. In these regions, the spectrum becomes highly sensitive to frequency changes, leading to rapid spectral variations in a narrow bandwidth. As a result, slight prediction errors can accumulate quickly, significantly lowering the $R^2$ value. However, in the frequency range between 10 GHz to 15.8 GHz, the prediction accuracy remains consistently high. That is because in this frequency band, the permeability of the YIG layers becomes negative, resulting in a stopband of RCP mode with vanished transmission and total reflection. Since the outputs of the deep learning models are the spectral differences, the learning task is effectively simplified in this region, resulting in high prediction accuracy. To quantitatively validate this explanation, we compute the mean absolute derivative of model's spectral response as a function of frequency, averaged over all samples and denoted by the green dashed line in Fig. 5. As seen, there exists a strong correlation between the derivative magnitude and the prediction accuracy: the frequency regions with higher derivatives tend to exhibit lower $R^2$ values, and vice versa.

To further understand how individual design parameters influence the band-limited nonreciprocal performance, we investigate the sensitivity and robustness of the designs generated by the VAE. The sensitivity here refers to how strongly the spectrum changes when a given parameter deviates from its original value, whereas robustness indicates the tolerance of the spectrum to such deviations. We generate a large number of candidate structures using VAE. The variations of the design parameters for these structures can be characterized by $\Delta_p = \dfrac{p_{f,\max} - p_{f,\min}}{p_{\max} - p_{\min}}$, where $p_{f,\max}$ and $p_{f,\min}$ denote the maximum and minimum values of the parameter $p$ within the feasible design set generated by VAE; $p_{\max}$ and $p_{\min}$ are the preset bounds of the parameter space. A smaller value of $\Delta_p$ indicates that the parameter $p$ is more tightly constrained by the band-limited requirement and therefore more sensitive.

While the examples in Fig. 4 focus on demonstrating the existence of feasible designs under stringent constraints, here we slightly relax the band-limited condition to obtain a sufficiently large set of feasible designs for statistical analysis. The results are summarized in Table 1, which lists the target frequency bands, the threshold $\eta$ for differential transmittance, the number of feasible designs $N$, and the normalized parameter variation $\Delta_p$ for all the design parameters $[d_1, d_2, d_3, d_4, d_5, \varepsilon_1, \varepsilon_2]$. As shown in Table 1, the smallest values of $\Delta_p$ consistently occur for the thickness of certain YIG layers (i.e., $d_3$ or $d_5$) indicating that the nonreciprocal performance is more sensitive to the geometric variations of these YIG layers, while the material parameters of the dielectric layers tend to exhibit greater robustness. While this property is helpful for practical realization of the structure, it is not general and may change for the structures with different number of layers.

| Band (GHz) | $\eta$ | $N$ | $d_1$ | $d_2$ | $d_3$ | $d_4$ | $d_5$ | $\varepsilon_1$ | $\varepsilon_2$ |
|---|---|---|---|---|---|---|---|---|---|
| 12~13 | 0.8 | 79 | 0.72 | 0.60 | **0.44** | 0.64 | 0.77 | 0.55 | 0.58 |
| 13~14 | 0.8 | 124 | 0.76 | 0.72 | **0.52** | 0.71 | 0.74 | 0.66 | 0.64 |
| 12~14 | 0.7 | 61 | 0.72 | 0.58 | **0.50** | 0.56 | 0.76 | 0.60 | 0.52 |
| 11~13 | 0.7 | 232 | 0.73 | 0.74 | 0.48 | 0.70 | **0.19** | 0.63 | 0.68 |
| 11~14 | 0.6 | 52 | 0.71 | 0.55 | 0.74 | 0.62 | **0.16** | 0.60 | 0.47 |

Table 1. Variations of the design parameters under different conditions.

To better understand the influence of the above parameter variations on the differential transmittance, we consider the case corresponding to the first row in Table 1 and compare the differential transmittance under perturbations of $\varepsilon_2$ and $d_3$. As shown in Fig. 6(a), varying the dielectric constant $\varepsilon_2$ by ±15% leads to only mild spectral changes, with the band-limited performance largely preserved. In contrast, Fig. 6(b) shows that the variations of thickness $d_3$ by ±15% result in significant spectral changes that deteriorate the nonreciprocal performance. In addition, we plot the electric-field amplitude distributions for the original structure [Fig. 6(c)] and the structures with 15% increase of $\varepsilon_2$ [Fig. 6(d)] and 15% increase of $d_3$ [Fig. 6(e)]. As see, all the three structures exhibit clear Fabry-Perot resonances arising from multiple reflections and transmissions. The field distribution in Fig. 6(d) remains similar to the original one in Fig. 6(c), while the field distribution in Fig. 6(e) shows a noticeable change. This is consistent with the results in Table 1, where the nonreciprocal performance of the system is more sensitive to the perturbations of $d_3$.

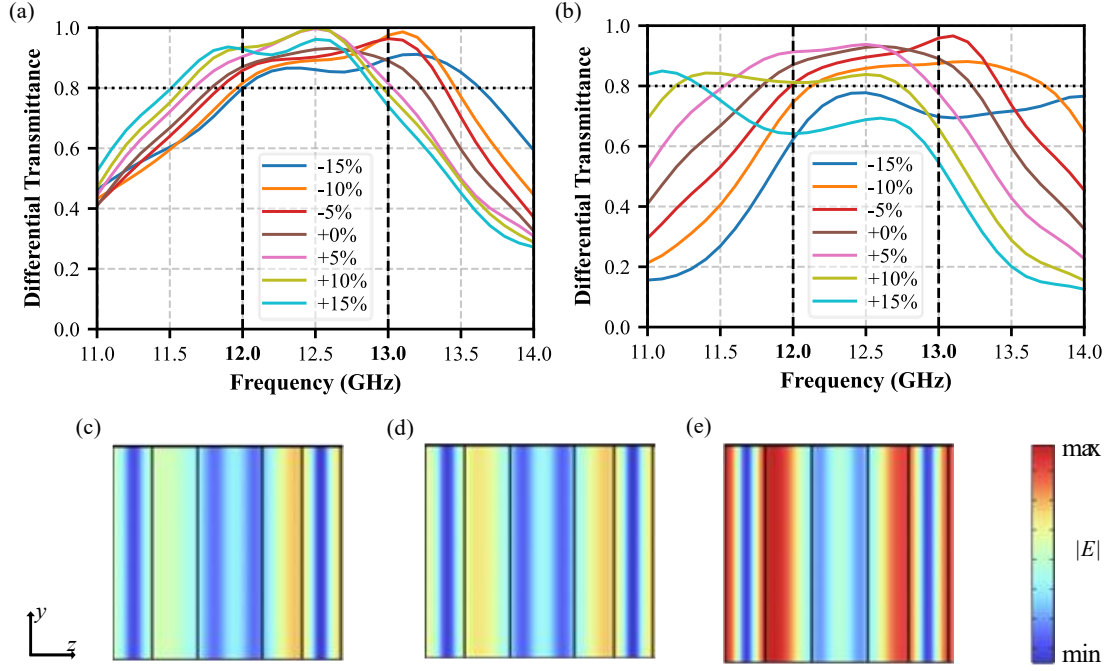

Fig.6. Band-limited differential transmittance spectra under different perturbations of (a) dielectric $\varepsilon_2$ and (b) thickness $d_3$. (a) Dielectric constant $\varepsilon_2$ is varied by ±15% around its ground value. (b) Thickness $d_3$ is varied by ±15% around its ground value. The vertical dashed lines indicate the boundaries of the target frequency band, and the horizontal dotted line denotes the desired threshold $\eta$ of differential transmittance. Distribution of absolute electric-field amplitude for the structures with (c) original design parameters [2.75, 3.18, 4.51, 2.78, 2.71, 3.92, 4.43], (d) $\varepsilon_2$ increased by 15%, and (e) $d_3$ increased by 15%. The frequency is fixed at 12.5 GHz.

## 7. Conclusion

In summary, we employ deep learning neural networks, specifically a forward prediction network, an inverse design network, and a variational autoencoder, to design and investigate nonreciprocal multilayer photonic structures. We use the dataset generated through TMM to train the three models, which exhibit strong predictive capabilities across the considered frequency spectra. Furthermore, by calculating the coefficient of determination and the absolute derivative, we identify a notable correlation between prediction accuracy and the system's intrinsic complexity. A quantitative sensitivity and robustness analysis of the design space reveals that the nonreciprocal performance of the structure is more sensitive to variations in certain structural parameters. These results indicate that deep learning models are effective not only for designing optical structures but also for probing the physical properties of nonreciprocal systems. Our work demonstrates the capabilities of deep learning in modeling and designing nonreciprocal photonic structures, facilitating the development of novel nonreciprocal devices for next-generation optical communications.

**Acknowledgement**

The work described in this paper was supported by grants from the Research Grants Council of the Hong Kong Special Administrative Region, China (Projects No. AoE/P-502/20 and No. CityU 11308223) and National Natural Science Foundation of China (No. 12322416).